\documentstyle[aps,epsfig]{revtex}
\begin{document}
\title{A study of the scenario with 
nearly degenerate Majorana neutrinos}
\author{Francesco Vissani}
\address{International Centre for Theoretical Physics,\\
Strada Costiera 11, I-34013 Trieste, Italy}
\date{August 28, 1997}
\maketitle
\begin{abstract}
Motivated by cosmological considerations, and
by the atmospheric and 
solar neutrino flux deficits,
we consider the scenario in which 
the three Standard Model neutrinos 
are endowed 
with a nearly equal 
Majorana mass in the 
eV range.
Combined constraints coming from 
(1) direct search for electron neutrino mass, 
(2) absence of neutrinoless 
double beta decay, and 
(3) unsuppressed electron neutrino 
flux in present reactor experiments, 
imply a quite 
well specified pattern of 
neutrino masses and mixing angles.
We discuss the experimental 
tests of the model and
comment on the features of this scenario.
\end{abstract}
\vskip0.4truecm
\noindent 
\section{Nearly degenerate Majorana neutrinos}
\subsection{Phenomenological character 
of the approach}
The possibility that the neutrinos
$\nu_e,$ $\nu_\mu$ and $\nu_\tau$ produced in weak interactions
are coherent combinations of almost degenerate
mass eigenstates  has attracted great attention in the
particle physics community 
\cite{deg1,deg2,deg3,deg4,deg5,deg6,deg7,deg8,deg9,deg10,deg11,deg12}.
This is motivated by the fact that the mass splitting 
required to explain the atmospheric and the solar neutrino 
flux deficits in terms of neutrino oscillations are much
smaller than the mass required for cosmological purposes,
as will be
discussed below  
in this first section. 

Most of the works focused on the
{\em ansatze} for the neutrino mass matrix: 
${\cal M}=m_0\times 1\!\!\!\!\!\,1\ \ +$ small 
corrections, and on the study of the corresponding 
theoretical motivations and phenomenological implications.
Inevitably, this {\em ansatze} implies
a signal for neutrinoless double 
beta decay ($0\nu \beta \beta $) 
experiments if $m_0=$ few eV`s, 
of interest for cosmological 
considerations. 
Allowing a departure 
from this mass pattern, the authors of \cite{deg9}
outlined a scenario in which the neutrinos have masses 
in the sub-eV range and could produce a signal 
in next generation searches of $0\nu\beta\beta.$ 
It was later proven that it is 
possible \cite{deg11} to incorporate a 
massive neutrino in the eV range
in the context of models 
with abelian family symmetry, in a way marginally consistent  
with the $0\nu \beta \beta $ bounds.

In the present work we do not commit 
{\em a priori} ourselves to any given 
pattern for the neutrino mass matrix,  and concentrate 
on finding a picture of the neutrino masses
and mixing angles that is consistent with the phenomenological
observations, and with the hypothesis of massive neutrinos
in the cosmologically interesting 
range. We will show that these
hypotheses are quite restrictive, 
implying that the corresponding scenario 
is likely to be confirmed or disproved by
forthcoming experiments.   

In its phenomenological character,
our approach is very similar to the one in \cite{deg12}. 
We emphasize in the present study 
the importance of the direct 
search for neutrino masses, as crucial 
test of the scenario under discussion. 
Another difference with \cite{deg12} 
is in the fact that the authors of that work concentrate 
on the MSW solution  
of the solar neutrino problem \cite{MS1,MS2,W}, 
whereas we focus on the vacuum oscillation solution \cite{BP};
in this respect, the two analysis complement each other.

\subsection{Definition of the scenario}
The lagrangian of neutrino masses in the basis where the 
charged lepton masses are diagonal is (bispinorial notation):
\begin{equation}
{\cal L}_{\rm mass}=-\!\!\sum_{\ell\ {\rm and}\ \ell'=e,\mu,\tau} \!
{\cal M}_{\ell \ell'}\ N_\ell(x)  N_{\ell'}(x)  + {\rm h.c.};
\end{equation}
the Lorentz indices are contracted according to $N^A N_A,$ $A=1,2.$ 
The symmetric matrix ${\cal M}$ can be decomposed as:
\begin{equation}
{\cal M}_{\ell \ell'}= \sum_{i=1,2,3} U^*_{\ell i}\  U^*_{\ell'i}\ m_{\nu_i} 
\end{equation}
(in compact notation: ${\cal M}=U^*\, m\, U^\dagger$), 
where $U$ is the unitary  matrix of mixing angles
and phases.
The three nonnegative and nondecreasing real numbers
$m_{\nu_i}$  are the Majorana neutrino masses. 
Introducing the bispinors $N_i(x)$:
\begin{equation}
N_\ell(x)=\sum_{i=1,2,3} U_{\ell i}\, N_i(x)\ \ \ \ \ \ \ell=e,\mu,\tau
\end{equation}
and the Majorana four-spinor in the chiral representation:  
$\nu_{i\alpha}(x)=(N_{iA}(x), \bar{N}^{\dot A}_i(x) )$
we can finally rewrite the mass lagrangian in the standard form:
\begin{equation}
{\cal L}_{\rm mass}=-\frac{1}{2} \sum_{i=1,2,3} m_{\nu_i} \,
\overline{\nu_i}(x)\ \nu_i(x) . 
\end{equation}
The mass difference is denoted as:
\begin{equation}
\Delta m^2_{ij}=m^2_{\nu_i}-m^2_{\nu_j} .
\end{equation}
It enters for instance in the expression 
$P_{\ell\ell'}=|\sum_i U_{\ell i}^*  U_{\ell' i} \exp(-i E_i t)|^2$ 
of the probability of oscillation in vacuum from a flavour
$\ell$ to $\ell'.$
In fact we can replace: 
$\exp(-i E_i t)/\exp(-i E_j t)$ by $\exp(-i\Delta m^2_{ij} t/(2 p)) ,$
under the assumption		
(which holds in the cases of interest) 
that the neutrino momentum $p$ 
is much larger than the neutrino masses.

The scenario under discussion can now be 
defined by letting in first approximation:
\begin{equation}
m_{\nu_i}= m_0\ \ \ \ \ \ i=1,2,3 .
\label{majorhp}
\end{equation}
The value of $m_0$ is fixed below,
with the help of cosmological 
considerations.
Before passing to that, we further specify two sub-cases
of this scenario, by making the assumption that 
the squared mass differences are
in ranges of interest to the solar and 
the atmospheric neutrino problems.
Case A is defined by:
\begin{equation}
\Delta m_{21}^2=\Delta m_{\odot}^2\approx 10^{-10} {\rm eV}^2 ,
\ \ \ 
\Delta m_{32}^2=\Delta m_{atm}^2\approx 10^{-2} {\rm eV}^2 ;
\label{figs}
\end{equation}
in case B the two mass differences are inverted
(compare with fig.\ 1 in last section).  
Since, apart from this fact (and neglecting
matter effects on the 
propagation of neutrinos), 
the two cases are phenomenologically 
equivalent, we will concentrate 
the discussion on case A.

\subsection{Neutrino masses of cosmological interest}
There are various indications of 
the existence of dark matter (DM---see for instance 
\cite{DMPDG}), 
that may constitute
the main component 
of the Universe.
Once the amount of dark matter, ordinary matter
and, possible other components are specified,
one can attempt a comparison of the observed 
structure with the one predicted 
in terms of the spectrum 
of the primordial fluctuations.
The gross features of 
the Universe can be explained by the
cold DM model, that assumes (1) that the
dominant component of DM is cold (non relativistic
at the moment of decoupling),
(2) a nearly scale invariant (Harrison-Zel'dovich) 
primordial spectrum, (3) and that the parameter $\Omega$
is 1, which implies that  
the total energy density is equal  to the critical density $\rho_c$
(the latter 
two hypothesis are suggested by inflationary scenarios). 
However COBE data have indicated 
the need of a certain
departure from this 
simple cosmological model \cite{hcdm}. 
The hot+cold DM model \cite{hcdm,hcdm2} 
achieves an optimal agreement
with the data if the hot DM constitutes 
the $20-30$\% of the energy density, 
the cold one the $70-60$\%, and the baryonic 
matter approximatively 1 tenth of 
the total energy density\footnote{In this model, 
the favoured value of the Hubble constant 
$H_0=h\times$ 100 km sec$^{-1}$ Mpc$^{-1}$ 
turns out to be around $h=0.5.$ 
In  \cite{HPDG} the observational 
range $h=0.65-0.85$ is suggested; a more 
recent analysis \cite{HST} gives 
$h=0.6-0.8$ as the reference range. 
Even lower values, $h=0.5-0.7$ have been 
advocated in \cite{prim}.}. 

Massive neutrinos are natural candidates for hot DM,
and this is the assumption we are interested to consider. 
The number density of neutrinos $n_\nu$ 
can be computed 
in the Big-Bang model \cite{wein}.
Once they become non-relativistic,  
their energy density reduces simply to
$\rho_\nu=\sum_i m_{\nu_i} n_\nu.$  
Assuming $\Omega=1,$ and
introducing the ratio of $\rho_\nu$ to the 
critical density $\rho_c$ (proportional to $ H^2_0$) 
one arrives at the 
expression that relates the neutrino masses to 
the cosmological parameters: 
\begin{equation}
\sum_{i} m_{\nu_i}=6.6\ {\rm eV} 
\times \left[ \frac{\rho_\nu}{0.2\, \rho_c} \right] 
\times \left[ \frac{h}{0.6} \right]^2. 
\label{cosm}
\end{equation}
The neutrino mass differences implied   
by solar and atmospheric neutrino deficits (\ref{figs}) 
are small in comparison with 
the mass in eq.\ (\ref{cosm}), 
and this facts suggest 
the hypothesis of nearly degenerate neutrinos.
Using eqs.\ (\ref{majorhp}) and (\ref{cosm}) 
together with the informations 
from the study of the structure evolution,  
we are lead to the following reference value for
$m_0=\sum_{i} m_{\nu_i}/3:$ 
\begin{equation}
m_0=m({\rm cosm.})=2.2\ {\rm eV} .
\label{preferred}
\label{figm}
\end{equation}
This corresponds to $h=0.6$ and $\rho_\nu=0.2\ \rho_c,$
or, with approximation
of 5\%, to $h=0.5$ 
and $\rho_\nu=0.3\ \rho_c.$
Progress in the determination of $H_0$ and $\Omega$  
will provide tests of the hot+cold DM scenario,
and also, if the scenario will be confirmed, 
a better determination of the parameter
$m_0.$

We briefly comment on the possibility that 
massive neutrinos may 
constitute the major component of the DM. 
It was suggested that this
could happen if special seeds of primordial 
fluctuations like those originated by the 
cosmic strings \cite{as} play a major role in the process of
structure evolution, 
or, alternatively, if neutrinos participate in 
exotic long-range (nongravitational) interactions 
\cite{rs}.
However, apart from any theoretical justification,
if this alternative scenario is correct,
and $\Omega=1,$
the reference value of the parameter $m_0$ would be 
larger than 10 eV, since $\rho_\nu$ in eq.\ (\ref{cosm})
increases almost $5$ times.

Of course, it is a major assumption to consider 
the absence of light sterile 
neutrinos, that could play various roles, and 
in particular could give the dominant contribution 
to the hot DM content of the Universe 
\cite{sterile1,sterile2,sterile3}.   
The hypothesis of (Standard Model) 
singlet neutrinos 
could be tested if they are mixed
with the ordinary ones, and 
participate in the oscillations. 
Regarding this point, 
if neutrino masses 
are the explanation not only of the solar and of the 
atmospheric neutrino problems, but {\em also} of the  
LSND data on $\bar{\nu}_\mu-\bar{\nu}_e$ 
\cite{LSND1,LSND1.5} and on ${\nu}_\mu-{\nu}_e$ 
flux conversion \cite{LSND2}, major modifications of the scenario, 
like the introduction of a sterile neutrino would be required. 
Therefore, in order to check the scenario
under discussion
(in which no light sterile neutrino is present
and a negligible signal at LSND is
expected) it will be crucial to see if the LSND findings will be 
confirmed by the forthcoming upgraded KARMEN 
\cite{KARMEN} and/or by other
experiments.

\section{Constraints from phenomenology}
In this section we discuss the bounds on the
neutrino masses and mixing angles that can be deduced from
the absence (1) of positive indications of electron neutrino mass
in the tritium decay experiments; (2) of neutrinoless double beta decay 
signal; (3) of appreciable disappearance 
of electron neutrino flux in
reactor experiments.
The constraints are considered in order of relevance for 
the scenario with three almost 
degenerate neutrinos\footnote{The importance 
of the direct searches for neutrino
mass has been not, in our opinion, sufficiently 
emphasized in the literature, even if the existence 
of this constraint was realized since the beginning
(see for instance fig.\ 2 in reference \cite{deg2}).}.
 
If one (or more) of the related experiments finds  
a signal, one (or more) parameters of the model will 
be determined. 
We stress that, in the model under discussion, 
$(i)$ the direct detection of neutrino mass seems to be possible,
and that $(ii)$ condition (2) implies a strong restriction
on the mixing angles, that will further strengthen if future 
$0\nu\beta\beta$ experiments will continue to give a null result.

\subsection{Direct search for neutrino masses} 
Let us denote by $d\Gamma[m_{\nu_i}]/d E_e$ the differential electron spectrum
in beta decay supposing that $\nu_i$ coincides with the electron neutrino.
In the general 
case, in which the electron neutrino is not
a mass eigenstate, the beta spectrum $d\Gamma/d E_e$ 
is the weighted combination:  
\begin{equation}
\frac{d \Gamma}{d E_e}=
\sum_{i=1,2,3} 
\frac{d\Gamma[m_{\nu_i}]}{d E_e}\ |U_{ei}|^2\ 
\Theta(\Delta M -m_{\nu_i}-E_e)  ,
\end{equation}
where the Heaveside function $\Theta(x)$
enters the formula due to kinematical requirements, 
and $\Delta M$ is the difference between the mass of the 
decaying nucleus and that of the final nucleus.
In our case, due to (\ref{majorhp}),
the unitarity relation
$\sum_i | U_{ei}|^2=1$ implies a constraint 
straightforwardly on the common mass $m_0.$  

Stringent limits can be  
derived by the experimental study of endpoint spectrum in 
tritium decay, after taking into account in some way 
the problematic indication of negative masses squared 
common to most recent experiments (corresponding to the presence of a 
{\em bump} in the endpoint region).
In the most recent experimental analysis \cite{Lobashev}, still under progress,
new parameters unrelated to the neutrino mass are introduced\footnote{In 
last analysis these parameters have be related to instrumental 
or maybe new physics effects.}. The upper limit obtained,
$m_\nu^2=1.5\pm 
5.9(\rm stat.)\pm 3.6(\rm syst.)$ eV$^2,$ 
implies: 
\begin{equation}
m_0<m({\rm {}^3 H})=3.9\  {\rm eV\ at\ 95 \%\ 
CL},
\label{trit}
\end{equation}
summing the errors in quadrature.
The limit (\ref{trit}) is not very different 
from those obtained by other experiments;
it could be further strengthened to 2.2 eV if 
the bump would be fully understood \cite{Lobashev}.

The limits on neutrino mass from SN1987A obtained
by the measured spread in time (which is related to the
original spread in energy and to the duration of the
neutrino pulse) are significatively looser,
$m$(SN1987A)
${\ \raisebox{-.4ex}{\rlap{$\sim$}} \raisebox{.4ex}{$<$}\ } 15$ eV. 
However, the observation of neutrinos 
from a very distant supernova 
could yield a competitive or even 
better bound in future.

The following 
remarks should clearly illustrate that, 
in the scenario with nearly degenerate neutrinos,
the direct search for neutrino mass
is of cosmological relevance:
\begin{enumerate}
\item The ``preferred'' value (\ref{preferred})
is quite close to the limit (\ref{trit}): 
$m_0\approx 2\times m({}^3{\rm H}).$
Therefore, the hot+cold DM model 
can be ruled out or confirmed by a slight increase 
in the precision of the measurements. 
We stress once more the importance of a clear 
theoretical description of the structure 
of the endpoint spectrum in tritium decay studies.
\item For $\rho_\nu=0.2\, \rho_c$ 
the limit (\ref{trit}) disfavours 
$h>0.8$ in the hot+cold DM model. 
Increasingly precise measurements 
of the cosmological paramenters and numerical simulations 
will be important in narrowing the parameter space of the model. 
\item The scenarios in which  
$\rho_\nu$ dominates the 
energy density of Universe 
and that have $\Omega=1$
require 
$m_0 {\ \raisebox{-.4ex}{\rlap{$\sim$}} \raisebox{.4ex}{$>$}\ } 10$ eV,  
and therefore can 
already be excluded in the present context.
\end{enumerate}

\subsection{Neutrinoless double beta decay and large mixing} 

The limit (\ref{trit}) (the preferred value (\ref{cosm})) are 
approximatively 6 times (3 times) larger than the limit 
obtained by searches of neutrinoless double beta decay \cite{ndbd}: 
\begin{equation}
\left|{\cal M}_{ee} \right|=
\left|\sum_{i=1,2,3} U_{ei}^2\, m_{\nu_i} \right|
<m(0\nu\beta\beta)
=0.68\ {\rm eV}\ \ \ {\rm at\ 95\%\ CL} .   
\label{nnn}
\end{equation}
We disregard in the following the possibility of large errors
in the estimation of the hadronic matrix elements.
The bound  (\ref{nnn}) implies  a cancellation 
in the contributions to $|{\cal M}_{ee}|$  
with a precision of 
$m(0\nu\beta\beta)/ m_0\approx 30$ \%. 
Let us first analyze the case in which the neutrinoless 
double beta decay is strongly suppressed, which implies
$\sum_i U_{ei}^2\approx 0.$ 
This equation can be represented geometrically as
a triangle in the complex plane. 
Therefore it can be satisfied only if 
the sum of the length of two sides is not smaller than   
the length of the third side. Taking into account 
that unitarity fixes to unity the perimeter of the
triangle, we conclude that the values
of the moduli for which 
cancellation is possible must satisfy:
\begin{equation}
\begin{array}{l}
\displaystyle
\frac{1}{2} \ge |U_{e1}|^2 , \\[2ex]
\displaystyle
\frac{1}{2} \ge   |U_{e2}|^2 \ge \frac{1}{2}-|U_{e1}|^2 .
\label{db}
\end{array}
\end{equation}
Assuming instead that a signal is observed,
the upper bound on the elements of the
mixing matrix resulting 
from geometrical considerations 
reads:
\begin{equation}
\left| U_{e j} \right|^2 \le \frac{1+ |{\cal M}_{ee} |/ m_0}{2} .
\label{bbb}
\end{equation}
Even with a maximal value of the mass ratio:
$|{\cal M}_{ee} |/ m_0=30$ \%
(that would imply that a signal of 
$0\nu\beta\beta$ 
will be observed close 
to the present limit) 
the upper bound does not change much: 
$|U_{ej}|^2\le 0.65.$ 
 
We want to make one point clear: we are not  proposing 
any explanations of the cancellation  between the 
elements of the mixing matrix squared,  we are just remarking 
that, once the hypotheses 
underlined above are accepted, consistency with the data
{\em requires a cancellation;}  this is true even 
if a positive signal, related to a small mass ${\cal M}_{ee}$ 
should be found. 
We show below that quite a precise 
pattern of the mixing matrix can be figured out
insisting on this scheme, and on the
consistency with the experimental data.

Eq.\ (\ref{db}) implies that there is at most 
one small mixing matrix element $U_{ej},$ 
and even in this case the other two 
elements correspond to maximal mixing angle.
The presence of large mixings, 
that was generally noticed in previous studies 
\cite{deg1,deg2,deg3,deg4,deg5,deg6,deg7,deg8,deg9,deg10,deg11,deg12} 
is simply a consequence of the absence of signal 
in $0\nu \beta\beta$ searches in these models.

It was remarked that the scenarios with large lepton mixing 
tend to be disfavoured by the comparison of 
the observed SN1987A neutrino energy spectrum 
and the theoretical spectra \cite{ssb}. 
Therefore, a clear understanding of the
mechanism of neutrinos emission and propagation
in the process of stellar collapse 
could provide us with the most direct way to exclude
the neutrino scenario under discussion.
We will not speculate further on this idea 
in the present work, and 
concentrate instead on the study of 
possible patterns of the mixing matrix.

\subsection{Information from the reactors} 
The neutrino beams produced at nuclear reactors 
are well suited for the study of disappearance 
of electron neutrino flux. 
The neutrino oscillations are 
described by the conversion probability
\begin{equation}
P_{ee}=1-4 \sum_{j>i} | U_{ej} |^2 | U_{ei} |^2 
\sin^2\left[ \frac{\Delta m_{ji}^2 L}{4\ p}\right] ,
\label{esurv}
\end{equation}
which just depend on the mixing angles which are
constrained by the neutrinoless double beta decay.
In the case A, since that the parameter 
$\Delta m_{\odot}^2$ 
is too small to develop oscillating phases 
on terrestrial distances, 
our expression (\ref{esurv}) reduces to:
\begin{equation}
P_{ee}=1- 4 | U_{e3} |^2 (1-| U_{e3} |^2)  
\sin^2\left[ \frac{\Delta m_{atm}^2 L}{4\ p}\right]\ \ \ \ ({\rm case\ A})
\label{www}
\end{equation}
that is effectively a two-flavour neutrino task, the 
relevant element of the
mixing matrix being the one related with the 
``isolated'' massive neutrino
(in case B, one arrive at 
the same expression with $U_{e3}$ replaced by $U_{e1}$).
The neutrino flux is not suppressed at all 
in the two extreme cases: $|U_{e3}|^2\approx 0$ or $\approx 1,$
but the second solution must be discarded 
in view of the absence of $0\nu\beta\beta$ signal, 
that implies eq.\ (\ref{bbb}). 
Applying the inequalities 
(\ref{db}) we conclude:
\begin{equation}
|U_{e1}|^2\approx 1/2,\ \ |U_{e2}|^2\approx 1/2,\ \  
|U_{e3}|^2\approx 0 \ \ \ \ ({\rm case\ A})
\label{sq}
\end{equation}
(in case B the small element is $|U_{e1}|$). 
The constraints from reactor experiments 
are not, of course, absolute. 
To be quantitative we notice  
that for a value of $\Delta m_{atm}^2=0.01$ eV$^2$ 
that may explain well the atmospheric anomaly
we get the bound \cite{krasn}: 
\begin{equation}
|U_{e3}|^2< 0.2 \ {\rm at\ 90\%\ CL}
\ \ \ \ ({\rm case\ A}).
\label{ubound}
\end{equation}
This bound however 
depends strongly on the 
value of $\Delta m_{atm}^2,$ 
and becomes already ineffective if instead of 
$0.01$ we have $7.5\times 10^{-3}$ eV$^2:$
This should warn from a literal use of equation (\ref{sq})
(compare also with the discussion in \cite{deg12}).
The future search at Palo Verde \cite{paloverde} and at
CHOOZ \cite{CHOOZ}) will be able 
to strengthen the bound, 
or maybe suggest a non-zero value of the mixing angle.
Assuming eq.\ (\ref{sq}), the  
bounds from neutrinoless double beta decay
are easily solved: $U_{e2}\approx \pm i\times U_{e1},$
where $i$ is the imaginary unity.  

\section{Features of the scenario}
Let us summarize the major consequences of the 
constraints exploited in previous 
section by using the matrix $|{\cal U}|,$ whose elements are
the numbers $|U_{ej}|.$
Using eq.\ (\ref{sq}), assuming for simplicity 
that $U_{e3}$ is quite small---our stronger 
hypothesis---, and the unitarity relations\footnote{In strict 
analogy with the what happens to the 
Cabibbo-Kobayashi-Maskawa mixing matrix, 
the matrix $|{\cal U}|$ 
can be described in the general case  
by four parameters (for instance one can use
three mixing angles and one phase).},
the approximate pattern is:
\begin{equation}
|{\cal U}|\approx 
\left(
\begin{array}{lll}
\frac{1}{\sqrt{2}} & \frac{1}{\sqrt{2}} & 0 \\
\frac{\sin\theta}{\sqrt{2}} & \frac{\sin\theta}{\sqrt{2}} & \cos\theta\\
\frac{\cos\theta}{\sqrt{2}} & \frac{\cos\theta}{\sqrt{2}} & \sin\theta
\end{array}
\right) \ \ \ \ ({\rm case\ A}) 
\label{matrixa}
\end{equation}
(in case B, the last column exchanges with the first one).  
There is  only one parameter in eq.\ (\ref{matrixa}),   
$\theta\in [0,\pi/2];$ 
we will see in the following discussion 
that the preferred value is $\theta\approx \pi/4.$ 
As discussed in sub-section C. above, 
the zero entry should be intended 
as an approximate statement only. 
In the rest of this work we will 
discuss the qualitative features 
of the scenario outlined, leaving  the 
detailed comparison with the data for
a future work.

\subsection{Atmospheric neutrinos}
On terrestrial distances $L\approx R_e,$
and for neutrino momenta $p\approx 1$ GeV the oscillations related 
to the mass difference $\Delta m_{\odot}$  
do not develop: $\cos(\Delta m_{\odot}^2 L/(2 p))\to 1.$  
Therefore, in case A, the expressions for the matrix ${\cal P}$
of the conversion probabilities (whose  elements  
are $P_{\ell \ell'},$ where $\ell$ 
and $\ell'=e,\mu,\tau$) is:
\begin{equation}
{\cal P} \approx 
1\!\!\!\!\!\,1 -\sin^2 2\theta\times 
\sin^2\left[ \frac{\Delta m^2_{atm}\, L}{4\, p}\right] \times 
\left(
\begin{array}{rrr}
0  & 0 & 0  \\
0 & 1  & -1 \\
0 & -1  & 1 
\end{array}
\right). 
\label{avep}
\end{equation}
If we suppose that the oscillations related to $\Delta m^2_{atm}$ 
are effectively averaged, at least at distances $L$
of the order of the Earth size, 
we can replace the oscillating factor 
$\sin^2(\Delta m^2_{atm} L/(4 p))$ by $1/2.$ 
A rough estimation of the muon-to-electron neutrino ratio 
$r=\Phi(\nu_\mu)/\Phi(\nu_e)$ measured in the detectors
can be obtained assuming that in the production region 
$r_0= \Phi_0(\nu_\mu)/\Phi_0(\nu_e)\approx 2,$ that 
is adequate for neutrinos of energies around 1 GeV
originated by cosmic rays entering the Earth atmosphere. 
Using $\displaystyle \Phi(\nu_\ell)=\sum_{\ell'=e,\mu,\tau} 
P_{\ell\ell'}\, \Phi_0(\nu_{\ell'})$ 
we find for the double-ratio $R:$
\begin{equation}
R\equiv \frac{r}{r_0}\approx 1 -\frac{\sin^2(2\theta)}{2} .
\end{equation} 
Kamiokande \cite{Katm} and Superkamiokande \cite{SKatm} measurements in
the multi-GeV energy region suggest $R\approx 0.55,$
and this implies that $\theta$ has to be close to
$\theta=\pi/4$ to produce a
large suppression factor.
However, the data are much better fitted 
if a linear dependence of the spectrum 
from the cosine of the zenith angle $\phi_Z$
is also introduced:
\begin{equation}
R(\phi_z)\approx 0.55+0.3\times \cos\phi_Z .
\end{equation}  
This favours not-so-large values of $\Delta m_{atm}^2,$ 
in such a way that the averaging effect is reduced
for smaller neutrino travel distance (around $10-100$ km, 
corresponding to $\phi_Z$ in a neighbourhood of 0) 
and it is maximal for neutrinos produced in the
antipodal regions (produced at
$L\approx 10000$ km far apart 
from the detector, 
that is for $\phi_Z\approx \pi$).
So the averaged oscillations would take place at  
maximal distances only: $R(\pi)\approx 0.3,$ which 
should be compared with the smallest possible 
theoretical value, $R\approx 0.5.$
This is not excluded by the data, the errors in
each of the $5\times 2$  data points being around 0.1
(except for $\phi_Z\approx 0,$ where the errors are $4-5$
times larger).
But while waiting for more precise data, one has 
to keep in mind that the model considered 
suggests somewhat smaller flux suppression 
than  measured at present.

\subsection{Solar neutrinos}
We consider the vacuum oscillation 
mechanism as a solution of 
the solar neutrino problem,
which requires $\Delta m_\odot^2\approx 10^{-10}$ eV$^2.$ 
The survival probability of electron neutrinos in case A is:
\begin{equation}
P_{ee}=1-2 |U_{e3}|^2 (1-|U_{e3}|^2)-4 |U_{e1}|^2 |U_{e2}|^2 
\sin^2\left[\frac{\Delta m_{\odot}^2 L}{4\ p}\right] 
\end{equation} 
where $L$ represents the Earth-Sun distance. 
The second term gives the oscillating factor, whereas 
the first term can give an additional (constant in energy) 
suppression. In fact, if $|U_{e1}|^2=|U_{e2}|^2$ we 
can rewrite:
\begin{equation}
P_{ee}=|U_{e3}|^4+(1-|U_{e3}|^2)^2 \cos^2\left[
\frac{\Delta m_{\odot}^2 L}{4\ p}\right]  .
\end{equation}
If the bound  (\ref{ubound}) holds, 
the first term is negligible; therefore the oscillating factor
can be multiplied by a suppression factor 
that can be as small as 0.64. 
This can somewhat diminish the spectrum distortion 
and the seasonal variations that are correlated 
characteristics of the vacuum-oscillation mechanism \cite{ms}.
If, beside that, we suppose that the oscillations are
effectively averaged (requiring  that
the splitting $\Delta m^2_\odot$ is 
large in comparison with 
$10^{-10}$ eV$^2$), the 
probability of suppression is close to the maximal 
{\em averaged} value $\langle P_{ee} \rangle_{max}=1/3.$ 
This may be of relevance for
neutrinos emitted by a supernova, or maybe 
for solar neutrinos themselves \cite{ap}.

\section{Summary and Discussion}
We summarize our findings in fig.\ 1, which
visualizes the  flavour content $|U_{\ell i}|^2$ of the
massive neutrinos $\nu_i$ in the way used in \cite{Smirnov}
to present a panorama of scenarios 
of neutrino masses and mixing.
\begin{figure}[htbp]
\centerline{\epsfig{file=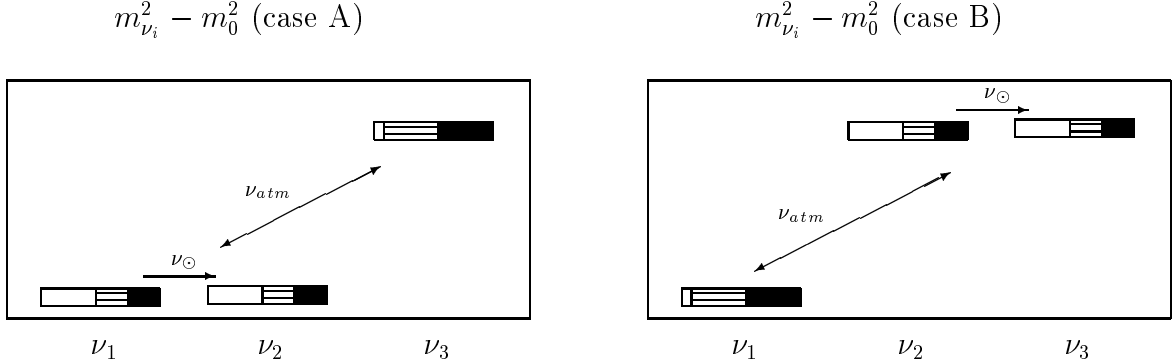}} 
\vskip1cm
\caption{Composition of massive neutrinos: 
white for electron-,  
hatched for muon- and black for tau-neutrino
flavour.
For a proper interpretation of 
the vertical scale, recall that 
the mass splittings, eqs.\ (\protect{\ref{figs}})
are much smaller than the common mass
scale $m_0,$ eq.\ (\protect{\ref{figm}}).}
\end{figure} 

Let us recall the main features of
the scenario proposed, that assumes that  
the solar and the atmospheric anomalies 
in the fluxes of neutrinos are
due to oscillations: 
\begin{enumerate}
\item Cosmological indications can 
be reconciled with the 
the solar and atmospheric neutrino deficits
if the three Standard Model neutrinos have 
in first approximation the same mass
$m_0=m({\rm cosm.}).$ 
\item We emphasized that
the common mass $m_0$ needed in hot+cold DM scenarios
can be probed 
in direct searches for neutrino masses, 
and that this mass is 
only {\em two times smaller} than 
the masses that can be currently excluded
(eqs.\ (\ref{preferred}) and (\ref{trit})).
The case in which the energy of the Universe is dominated
by three massive neutrinos can already be excluded.
\item The absence of a signal 
in neutrinoless double beta
decay searches requires a cancellation. Even if this is 
not {\em a priori} a theoretical attractive  
perspective,
it implies strong restrictions on certain 
combinations of the elements of the mixing matrix
(compare with eq.\ (\ref{db})).
\item The absence of signal in 
reactor experiments suggests that 
the structure of the mixing matrix,
eq.\ (\ref{matrixa}),  
may be quite simple
(see also eq.\ (\ref{ubound}) 
and related discussion).
Regarding this point, 
future electron neutrino 
disappearance experiments will be 
important to better fix the parameters of the model.
Even better, they could permit us to distinguish the present 
variant of the scenario with nearly degenerate neutrinos
with the variant in which the solar 
neutrino deficit is explained invoking the 
MSW effect; in fact, in this latter case 
{\em large} values of the mixing element $U_{e3}$
are preferred \cite{deg12} (compare 
with eq.\ (\ref{www}) and discussion thereafter).
  
\item The pattern of the mixing 
matrix allows
us to take into account the 
atmospheric neutrino anomaly. 
With accumulation of data at SuperKamiokande, 
the zenith angle dependence of the 
neutrino spectrum will become 
a severe test
of the scenario proposed.
\item Regarding the observed 
suppression of the solar neutrinos flux, 
we remarked that this model can incorporate
the vacuum oscillation solution, 
possibly 
with reduced distortion of the energy spectrum.
\end{enumerate} 

In conclusion, we express our opinion
that, even if the model proposed  is not 
at present of particular theoretical appeal,
it is of some interest since $(a)$ it incorporates 
cosmological and phenomenological indications;
$(b)$ it is enough restrictive to be 
confirmed or disproved in the future (few) years.

\acknowledgments{I am grateful to the Theory Group of 
FERMILAB, and in particular to the Deputy Head
of the High-Energy section 
R.K.\ Ellis, 
for hospitality during the last stages of this work.
It is also a pleasure to thank 
J.\ Peltoniemi (who created and maintains the database 
{\em The Ultimate Neutrino Page}
at the URL http://www.physics.helsinki.fi/neutrino/)
for scientific help and also for encouragement;
R.N.\ Mohapatra, 
M.\ Persic, 
A.\ Riotto, 
E.\ Roulet, 
P.\ Salucci and 
A.\ Stebbins
for discussions about aspects of cosmology;
E.Kh.\ Akhmedov and 
M.V.\ Chizhov 
for discussions about neutrino phenomenology.
I need special 
thanks for A.Yu.\ Sminorv, who introduced
me to the subject of neutrino physics and helped me
to orient in this vast field.}

\end{document}